# Random scattering and alternating projection optimization for active phase control of a laser beam array

**J. SAUCOURT**[1,2,*], **P. ARMAND**[1], **V. KERMENE**[1], **A. DESFARGES-BERTHELEMOT**[1], **AND A. BARTHELEMY**[1]

[1] *Université de Limoges, CNRS, XLIM, UMR 7252, 123 Avenue Albert Thomas, F-87000 Limoges, France*
[2] *CILAS, 8, avenue Buffon, F - 45063 Orléans, France*
*\*jeremy.saucourt@xlim.fr*

**Abstract:** A fraction of a laser beam array, whose unknown piston phase relationships must be set to prescribed values, is launched into a scattering media with random transmission. The resulting output speckle pattern is sampled by an array of photodiodes measuring the local light intensity. The data feed an innovative optimization process which controls a phase modulator array. Few iterations of the opto-numeric loop lead to efficient and fast phase locking on any desired piston phase distribution.

## 1. Introduction

Array of tiled coherent laser beams, either from a laser array or derived from a single laser oscillator and further individually amplified, is a common approach to increase the power delivered by laser sources [1,2]. Coherent combining of the individual laser beam raises in turn the source brightness opening an opportunity to satisfy in particular the need of future demanding applications such as fiber laser wake field acceleration of particles [3] or light propulsion of nanocraft in space [4]. Coherent laser field summation of a beam array requires free space propagation of the beams with a controlled phase distribution. It is well known that in-phase beams will interfere in the far field giving a maximum power on axis for example. This is the reason why many techniques have been developed to phase-lock the fields of a laser array [2]. For some other applications a fast non-mechanical beam steering is sought [6,7], or a beam forming capability or a compensation of the distortions due to atmospheric turbulence [8]. In those cases, the phase distribution in the array has to be no longer uniform. In the context of laser arrays, a tiled beam array produces a discretized wavefront that differs from the usual continuous wavefront of a beam with aberrations or distorted by its transmission through the atmosphere. Most of times, in this context, the wavefront to control takes the form of a discrete distribution of phase steps (piston-phase). Some phase-locking techniques have the capability to shape any desired discrete wavefront but they require a fast measurement of the instantaneous individual phase of the whole beams of the emitter array either with help of a reference plane wave [9] or with shearing interferometry method [10]. Beam pointing on a distant target through turbulent atmosphere requires also the proper adjustment of the phase pattern in the beam array launched toward the target. That was achieved with target-in-the-loop devices based on an adaptive SPGD approach [11] as well as by a frequency tagging technique (LOCSET) [12], both using a single detector close to the laser source. The target was cooperative enough to give a feedback signal of sufficient level. In this paper we will not consider a target-in-the-loop scheme but we will assume that the phase chart desired for the delivered beam array is known. We present an alternative approach for piston phase control of a tiled beam array. It is based on an iterative method including the transformation of the beam array into a speckle pattern by a scattering plate and its combination with a specific optimization process. The proposed approach shares some common features with the Phase-Intensity Mapping (PIM) method [13-14], based on a phase contrast imaging device whose data served for an iterative optimization algorithm. In contrast to the latter, the new technique permits to phase-lock the beams on any arbitrary phase distribution across the array, and it offers a great flexibility for a compact implementation, even for arrays of large size.

## 2. Principle of the method

Let us first consider an array of n input laser fields of identical frequency that we denote by the vector $a \in \mathbb{C}^n$ whose phase distribution $arg(a)$ is unknown but where $|a|$ is known. The goal is to set the phase of vector $a$ on a desired set of values given by the vector $arg(d)$ with $|d| = |a|$. In the set-up a beam splitter picks up a weak fraction of the beam array which is sent onto a scattering plate. Scattering, which is an intrinsic random process, mixes the field from initially separated beams and creates multiple path interference giving a typical speckle pattern on a screen placed behind the plate at a distance that can be short. The intensity image of the speckle carries a nonlinear but deterministic encoding of the phase relationships in the input beam array. The situation is here different from the usual coherent diffraction imaging where the single input beam is known (a wide Gaussian laser beam with flat wavefront) and where scattering is due to the object whose shape must be determined. The proposed scheme is a kind of reverse case where the object is known (a diffuser) and it is the complex input laser field which is sought. In order to reduce the computation cost and to speed up the method, instead of the full image, an array of only a few photodiodes measure the speckle intensity in a reduced number m of points of the interference pattern. The m photodetectors yield an intensity vector from which one derives the amplitude vector $b \in \mathbb{R}_+^m$. It is connected to the input fields by the equality $|Xa| = b$, where $X \in \mathbb{C}^{m \times n}$ stands for the transmission matrix of the scattering device. To meet our objective, we need to find the vector of phase errors $e$, between the input fields $a$ and the desired fields $d$, such that:

$$|e| = 1 \text{ and } |X(d \cdot e)| = b \tag{1}$$

We want for each line $i$ of the vectors:

$$arg(a_i) - arg(e_i) = arg(d_i) + \delta \tag{2}$$

, $\delta$ being a constant phase offset within the range [0-2π]. So, the measured intensity values are processed to derive, from Eqs. 1 and 2, the phase modulation to apply on each beam in order to make the phase pattern in the beam array as close as possible to the desired distribution. In fact, a single correction round is not sufficient to lead to the desired shaping. But a few iterations of the same error correction process quickly converge and set the beam array on the desired phase pattern. The feedback loop can be kept in operation to preserve the shaping whatever the changes in the input fields.

The method we proposed to set the array phase on a desired distribution follows an iterative error reduction approach which different steps are summarized below.

---

**Opto-numerical loop for array phase setting on a desired distribution**

*Inputs:* Desired fields for the array $d \in \mathbb{C}^n$, scattering matrix $X \in \mathbb{C}^{m \times n}$

1. *Initialization*
   Step $k = 0$ and input vector $a = a_0$ with unknown initial state for the phase
2. *Optical scattering of the input fields and speckle amplitude measurement*
   Measurements give the amplitude $b_k$
3. *Inner algorithm for phase recovery*

    Compute approximate solution $a_k \in \mathbb{C}^n$ of the equation $|Xa_k| = b_k$
4. *Phase modulation of the input*
   Apply phase correction according to $\theta_k = arg(d) - arg(a_k)$
5. *Iteration*
   Iterate $k = k + 1$ and feedback to step *2*

---

A general schematic view of the principle is shown on Fig.1. The whole approach is equivalent to a phase retrieval procedure. However, generally speaking, standard phase retrieval is a fully computational treatment of a given set of experimental data. This is major difference with the present case, where at each optimization round, a phase modulation changes the input $a$ which in turn modifies the speckle pattern and the measured intensity $b^2$ (element-wise square). So, the data feeding the inner algorithm are updated at each round.

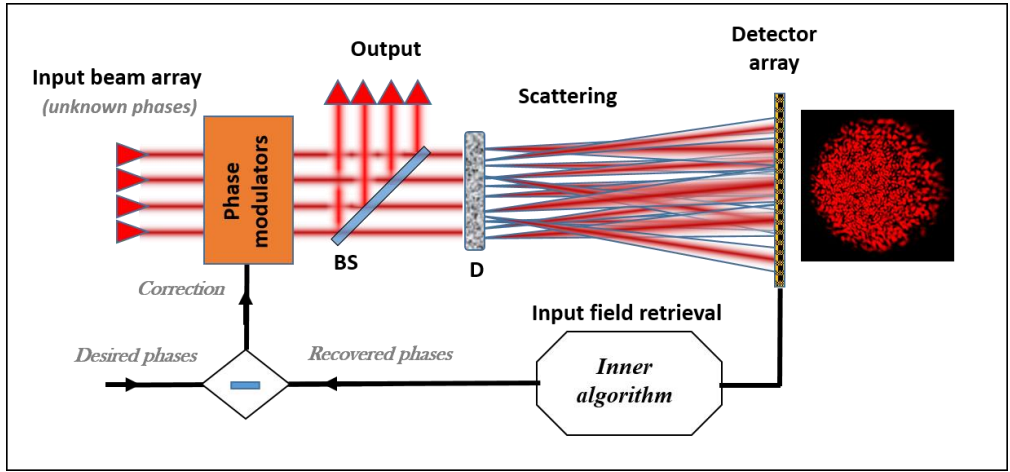

Fig.1: Schematic view of the principle for setting and maintaining the unknown phases of a laser beam array onto a desired transverse distribution.

In the following we will assume that the transmission matrix $X$ of the scattering device is known; its measurement will be reported in the next section. The estimated phases of the inputs $a$ are computed in a sub-part by a purely numerical iterative phase retrieval process which is based on alternating projection [15, 16]. The phase difference between the retrieved input field $a_K$ and the desired field $d$ indicates the phase modulation to apply on the modulators to get the desired discrete wavefront:

$$\theta_{corr} = arg(d) - arg(a_K) \qquad (3)$$

As mentioned above, a single correction round is not sufficient to lead to the desired shaping. This is due to the fact that the retrieved field is different from the actual input field. But a few iterations suffices to quickly converge on the desired phase pattern (see following section).

### 3. Numerical analysis

For a preliminary assessment of the proposed principle we performed numerical simulations. The beam array considered here was of square lattice with beams of identical power and size (uniform cross-section). Each beam was assigned a single complex value representing its amplitude and phase. The behavior of the scattering device was taken into account by a complex value random matrix. In order to mitigate the impact of the desired phase patterns, they were randomly chosen and changed at each simulation run. The phase distribution in the input beam array was also randomly varied for each run to consider the fact that in practice it will be unknown. The simulations results were then gathered and averaged for display and evaluation of the performances. The difference between the shaped wavefront and the desired discrete wavefront was assessed by a parameter named "phasing quality", noted $Q$, and calculated as the normalized scalar product of the two field distributions:

$$Q = \frac{|\langle a,d \rangle|^2}{\langle |a|,|d| \rangle^2} = \frac{\left|\sum_{i=1}^{i=n} a_i \cdot d_i^*\right|^2}{\left[\sum_{i=1}^{i=n} |a_i| \cdot |d_i^*|\right]^2} \quad (4)$$

A parameter $Q = 0.96$ corresponds to λ/30 rms deviation between the actual field and the target field [24]. We first study how many detectors may be sufficient to get the expected phase profile in order to minimize the computation cost and maximize the convergence speed. The numerical results demonstrate (see Fig.2) that the number of samples taken in the scattering pattern must be at least three times larger (four times is better) than the number of area to control in the input signal (number of beams in the array) to reach the same average phasing quality in steady state regime. We can find some similar criteria in the literature on optimization where it was reported that $(4n - 4)$ measurements are required to guarantee the uniqueness of a numerical phase retrieval problem (injectivity) on a complex vector of n elements [17]. Simulations show that convergence to the desired state is fast, taking on average less than five iterations of the process (five phase corrections).

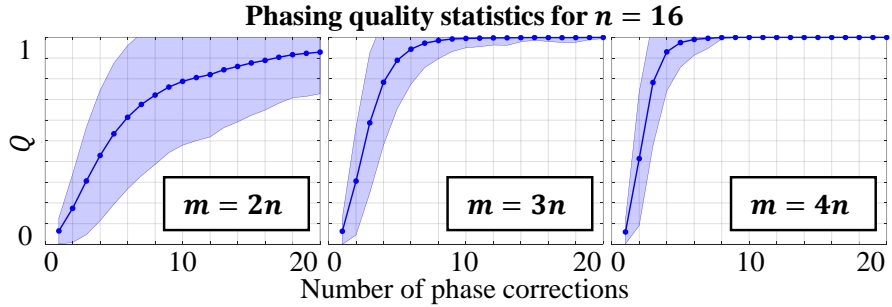

Fig.2: Average phasing quality $Q$ versus the number of phase corrections and for different ratios between the number of beams to control ($n$) and the number of detectors ($m$) in the speckle pattern. Statistics were computed from 1000 realizations. For each realization, the target phases vector and the initial phases vector were randomly chosen. Blue shaded areas denote standard deviations from average.

We study also how the number of iterations needed to reach the desired profile evolves with the number of beams to control. The simulations have shown that this number raises weakly (less than x2) when the number of waves increases by one order of magnitude (see Fig.3).

The impact of noise was also investigated. We have considered electrical noise in the measurement of optical intensity by the photodiodes and optical power fluctuations of the individual laser beams. Based on our simulations we concluded that up to 10 % rms noise on the photodiode signals and up to 5 % rms power wonder in the beam array have no visible impact on the efficiency of the phase adjustment as well as on the speed of the convergence to the desired discrete wavefront. The effect of such noises is also independent from the number of beams to control. All the observed behaviors attest that the technique is actually robust with respect to noise.

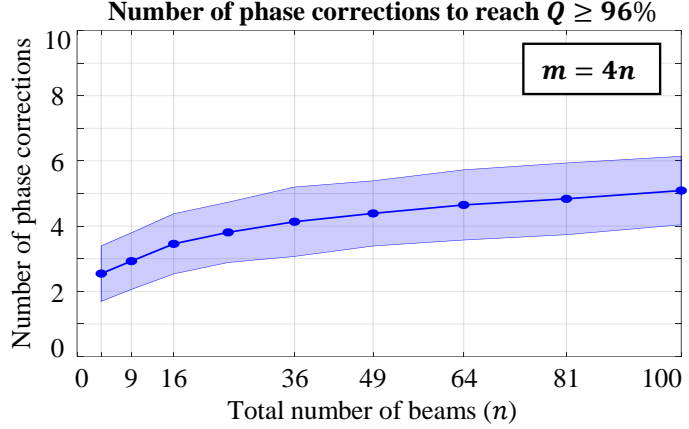

Fig.3: Average number of phase corrections (1000 realizations) to get phasing on the desired target, with less than λ/30 rms deviation ($Q \geq 96\%$), versus the total number of beams $n$ in the array. Blue shaded areas denote standard deviations from the average.

## 4. Scattering device transmission matrix measurement

A crucial point of the proposed approach is the precise knowledge of the complex value transmission matrix of the scattering device connecting the input laser fields to the output fields on the position where the photodiodes are located. There are several options to perform such a characterization which is frequently achieved in view of beam control and imaging through scattering media [18, 19]. The difficulty again comes from the fact that access to the phase is difficult in optics since photo-detectors only give the light intensity. In our case we chose a reference-less technique which is more convenient to implement [20]. It is based on series of intensity measurements $B \in \mathbb{R}^{N \times m}$ obtained from the m detectors for a sufficiently large set of input fields $A \in \mathbb{C}^{N \times n}$ made of $N$ successive random phase patterns (of size $n$, the number of beams) with uniform amplitudes. An optimization routine computes the elements $x_{i,j}$ of the transmission matrix $X \in \mathbb{C}^{m \times n}$ which are solutions of the unconstrained non-convex optimization problem:

$$\text{minimize } f(X) := \| |A.X^T|^2 - B \|^2, X \in \mathbb{C}^{m \times n} \tag{5}$$

There are many possible algorithms available to solve this problem. We have used here the same alternating projection iterative algorithm used for the phase retrieval algorithm above. Just we seek for the recovery of the transmission matrix $X$ instead of the recovery of the input vector $a$. We changed also the starting for a Wirtinger Flow initialization [21]. As an example, in the following experiments we have used as a scattering media a commercial ground glass diffuser (Thorlabs DG05-220-MD). An array of 16 laser beams at 1064 nm in a square lattice shine the random media sample. A spatial light modulator (Hamamatsu X131138) was inserted on the path between the laser source and the scattering plate in order to adjust the phase of the individual input beams. A camera was used to record the scattered intensity pattern in the far field, at a few centimeters distance from the diffuser. The intensity level on some selected pixels only was exploited in order to mimic a photodiode array of low size (8x8) but distributed on a wide surface of the light pattern. The intensity measurements corresponding to $N = 20 \times n$ random phase charts at input were sent to the optimization software which gave the experimental transmission matrix illustrated on Fig. 4.

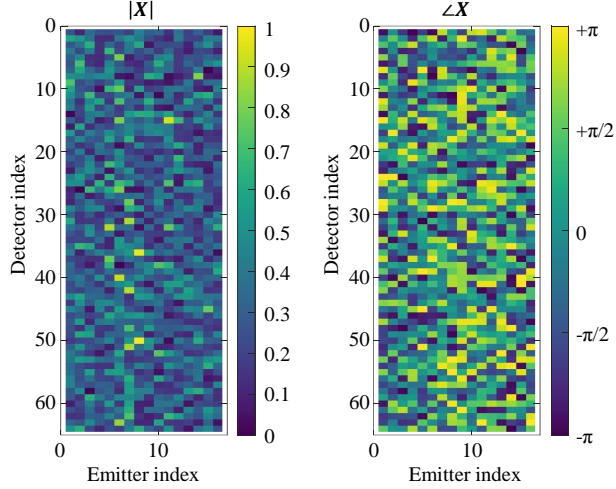

Fig. 4: Experimental scattering matrix (amplitude on the left hand side and phase (rad) on the right side) recovered from intensity measurements for a 4x4 input beam array and a 8x8 detector array.

Phase (on the right hand side) and modulus (on the left hand side) of the measured $X$ matrix looks mostly random as expected. Computation time was of the order of 10 ms (CPU Intel Core i5-8600K @3.60 GHz, Windows 10 64 bit, MATLAB R2018b) for this 64x16 matrix after recording of the experimental data. It was checked that the statistical properties of the measured matrix are similar to the ones of a random matrix. In particular we checked that the distribution of the singular values of the measured matrix is close to the Marcenko-Pastur distribution expected for a random matrix [22]. This ensures that the results derived from the previous numerical analysis based on a random matrix should be expected in the experiments. Once the transmission matrix of the scattering device has been characterized it was further used in the phase shaping experiment.

## 5. Proof of principle experiment

The experimental set-up used for a proof of principle experiment is schematically depicted in Fig. 5. The beam from a fiber coupled laser diode at 1064 nm (FC-LD) is first expanded by a magnifying telescope L1-L2 to cover the first reflective spatial light modulator (SLM1-Hamamatsu X 131138) surface. Then, a 4x4 laser beam array was generated by diffraction on an array of disk-shaped phase gratings displayed on SLM1. Beamlets were circular in shape with a diameter of 300µm and a 500µm period. In addition, the relative position of the gratings 'modulation was used to introduce a non-uniform phase piston among the beams in order to mimic unknown random initial phase distribution. The 4f telescope L4-L5, with 1:1 magnification, served to filter out the diffracted part of the SLM1 output thanks to the hard aperture A located in the focal plane of L4. A second spatial light modulator (SLM2, Hamamatsu X 10468) applied the phase corrections computed by the system to the individual beams. A beam splitter (BS) sent a fraction of the beam array on a diffuser (D) while the transmitted fraction was directed onto a positive lens (L6) for display of the far field pattern and its observation with a camera (CAM2-Thorlabs CMOS DCC1545M). The diffuser scattered the beams which formed a speckled figure on a camera (CAM1-Quantalux ScMOS CS2100M). In the detected images, 8x8 regions of interest were selected to mimic an array of photodiodes. Thus CAM1 provided 64 intensity measurements which were further processed by the algorithm implemented on a laptop computer. Operation of the whole set-up was interfaced.

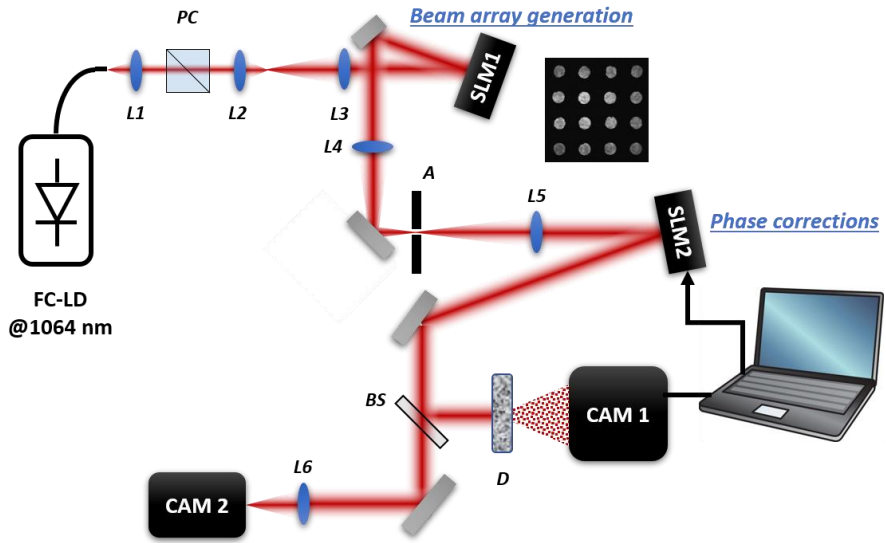

Fig. 5: Set-up for the proof of principle experiments.

The initial conditions (phase distribution in the array) as well as the desired phase charts were chosen with a random generator of uniform distribution in [-π, π].The figure 6 below reports the plot of the phasing dynamics according to the number of iterations of the opto-numeric optimization process. The data (red dots) were averaged on 100 phasing experiments where both the initial conditions and the phase targets were varied at each trial. The curve shows that 5-6 iterations only were sufficient to reach steady-state with a phasing quality of 96% (deviation from the target was less than λ/30 rms). It demonstrated the efficiency of the scattering-based optimization. The experimental dynamic is consistent with the one derived from our simulation ($m = 4n$) which is plot in blue solid line in Fig. 6. The total absence of noise in the modeling permits to reach an ideal final state with 100% phasing quality. Noise and above all the residual aberrations in the collimated laser diode beam, explain why the final phasing quality is slightly lower than 100% in the experiments. The red shaded area in the graph denotes the root mean square variation around the average and shows by experiments the robustness of the method.

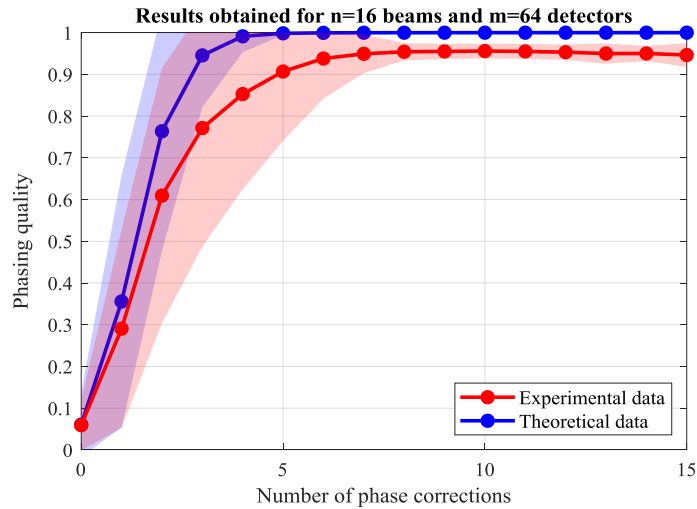

Fig. 6: Dynamics of the phasing quality on desired phase charts averaged on 100 realizations with varying initial conditions and targets. Measured data (red dots) and guide for the eyes (red solid line). For comparison, theoretical data from numerical simulation are displayed as blue dots connected by a blue solid line. The red and blue shaded areas show the standard deviations from the average in the experiments and in the simulations respectively.

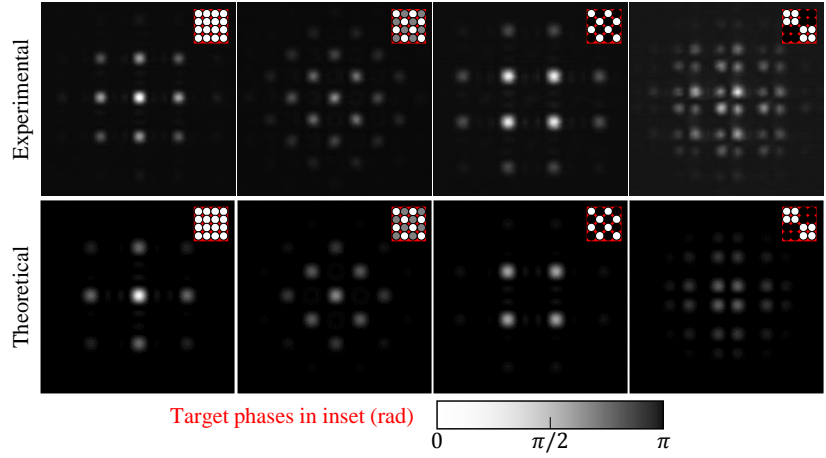

Fig. 7: Examples of experimental far field intensity pattern for a 4x4 laser beam array after adaptive shaping of their individual phase on the desired phase patterns given in inset. The initial phase conditions were unknown and randomly distributed. The insets represent the target phases. The scales of the experimental and theoretical far fields are slightly different.

The time spent on one iteration in these experiments (~200 ms) was mostly determined by the transmission of the command to SLM2 and above all by its time response. On the opposite, the computation of the phase corrections was fast and lasted less than 70 µs only. As examples, we present on Fig. 7 some experimental far fields of the beam array recorded after phasing on the desired phase patterns shown in insets, starting from unknown initial conditions. The far-fields observed after convergence were perfectly consistent with the theoretical expectations.

**7. Conclusion**

We presented in this letter a new optimization technique for setting the piston phase of a coherent laser beam array on any desired distribution. It is based on an optical scattering process which transforms the input beam array into a speckle pattern whose intensity is measured in a few transverse positions only. On the numerical side, alternating projection method is used to compute the phase modulation to apply in order to reach the desired phase profile, after few iterations. This is the first iterative optimization technique able to adjust the discrete wavefront of a laser beam array on a non-uniform distribution. It is the first time as well that a random scattering process is used for phase-intensity mapping and for phase control in the transverse cross section of a coherent beam. Numerical simulations with random scattering matrices validated the new technique and showed its efficiency. Five to nine rounds of phase correction only are sufficient on average to reach any desired phase chart for an array with up to 100 beams. A simple method is proposed to get the complex transmission matrix of a commercial scattering plate. It was further exploited in a proof of principle experiments with a 4x4 laser beam array and liquid crystal SLM for phase control. Starting from unknown initial conditions, phasing of the laser array was achieved on various phase charts with λ/30 rms accuracy after less than five iterations. This attests that the innovative method is efficient, fast and relevant for tuning and shaping the far field of a laser beam array with unknown and changing input conditions. The fact that the diffuser performs a linear random transform of the input fields enhance the efficiency of the phase retrieval and transmission matrix retrieval, as it was recently demonstrated [23]. The same approach might be used for phase recovery of a coherent image even in the case of strong intensity modulation or in the case of sparse intensity distribution.

**Funding**

Partial funding by CILAS Company (Ariane Group) under grant n °2016/0425.

**References**

1. T. Y. Fan, "Laser beam combining for high-power, high-radiance sources," IEEE J. Sel. Top. Quantum Electron. 11, 567–577 (2005).
2. A. Brignon Ed., *Coherent Laser Beam Combining* (Wiley-VCH, 2013).
3. G. Mourou, B. Brocklesby, T. Tajima, and J. Limpert, "The future is fibre accelerators," Nat. Photonics 7, 258 (2013).
4. Gilster, Paul (12 April 2016). "Breakthrough Starshot: Mission to Alpha Centauri". Centauri Dreams. https://www.centauri-dreams.org/2016/04/12/breakthrough-starshot-mission-to-alpha-centauri/
5. Benford, J., "Starship Sails Propelled by Cost-Optimized Directed Energy," Journal of the British Interplanetary Society, **66**, p. 85 (2013).
6. M. J. R. Heck, "Highly integrated optical phased arrays: photonic integrated circuits for optical beam shaping and beam steering," Nanophotonics **6**, 93–107 (2017)
7. M. J. Byrd, C. V. Poulton, M. Khandaker, E. Timurdogan, D. Vermeulen, and M. R. Watts, "Free-space Communication Links with Transmitting and Receiving Integrated Optical Phased Arrays," in Frontiers in Optics / Laser Science, OSA Technical Digest (Optical Society of America, 2018), paper FTu4E.1.
8. H. Bruesselbach, S. Wang, M. Minden, D. C. Jones, and M. Mangir, "Power-scalable phase-compensating fiber-array transceiver for laser communications through the atmosphere," J. Opt. Soc. Am. B **22**, 347–353 (2005).
9. H. Chosrowjan, H. Furuse, M. Fujita, Y. Izawa, J. Kawanaka, N. Miyanaga, K. Hamamoto and T. Yamada; "Interferometric phase-shift compensation technique for high power tiled aperture coherent beam combination," Opt. Lett. **38**(8), 1277-1279 (2013).
10. C. Bellanger, B. Toulon, J. Primot, L. Lombard, J. Bourderionnet, and A. Brignon, "Collective phase measurement of an array of fiber lasers by quadriwave lateral shearing interferometry for coherent beam combining," Opt. Lett. **35**(23), 3931-3933 (2010).
11. T. Weyrauch, M. Vorontsov, J. Mangano, V. Ovchinnikov, D. Bricker, E. Polnau, and A. Rostov, "Deep turbulence effects mitigation with coherent combining of 21 laser beams over 7 km" Opt. Lett. 41, 840–843 (2016)
12. P. Bourdon, V. Joliveta, B. Bennaia, L. Lombarda, G. Canata, E. Pourtala, Y. Jaouen, and O. Vasseur, "Coherent beam combining of fiber amplifier arrays and application to laser beam propagation through turbulent atmosphere," Proc. SPIE 6873, 687316 (2008).
13. D. Kabeya, Vincent Kermène, M. Fabert, J. Benoist, A. Desfarges−Berthelemot, A. Barthélémy, "Active coherent combining of laser beam arrays by means of phase-intensity mapping in an optimization loop", Opt. Express, **23**, 31059-31068 (2015).
14. D. Kabeya, V. Kermène, M. Fabert, J. Benoist, J. Saucourt, A. Desfarges-Berthelemot, A. Barthélémy, "Efficient phase-locking of 37 fiber amplifiers by phase-intensity mapping in an optimization loop," Opt. Express **25**, 13816-13821 (2017)
15. R. W. Gerchberg and W. O. Saxton. "Practical algorithm for determination of phase from image and diffraction plane pictures". Optik, 35(2) :237-246,1972.
16. P. Netrapalli, P. Jain, and S. Sanghavi. "Phase retrieval using alternating minimization". IEEE Trans. Signal Processing, 63(18) :4814-4826, 2015.
17. A. S. Bandeira, J. Cahill, D. G. Mixon, and A. A. Nelson, "Saving phase: Injectivity and stability for phase retrieval," Appl. Computat. Harm. Anal., vol. 37, 1, pp. 106−125, (2014).
18. S. M. Popoff, G. Lerosey, R. Carminati, M. Fink, A. C. Boccara, and S. Gigan, "Measuring the transmission matrix in optics: An approach to the study and control of light propagation in disordered media," Phys. Rev. Lett. **104**, 100601 (2010).
19. Y. Choi, T. D. Yang, C. Fang-Yen, P. Kang, K. J. Lee, R. R. Dasari, M. S. Feld, and W. Choi, "Overcoming the diffraction limit using multiple light scattering in a highly disordered medium," Phys. Rev. Lett. **107**, 023902 (2011).
20. A. Dremeau, A. Liutkus, D. Martina, O. Katz, C. Schulke, F. Krzakala, S. Gigan, and L. Daudet, "Reference-less measurement of the transmission matrix of a highly scattering material using a DMD and phase retrieval technique", Opt. Express 23 (9) 11898-11912 (2015).
21. E. Candes, X. Li, M; Soltanolkotabi, "Phase retrieval via Wirtinger Flow: Theory and algorithms" IEEE Trans. On Information Theory, 61 (4) 1985-2007 (2015).
22. V. A. Marcenko and L. A. Pastur, "Distribution of eigenvalues for some sets of random matrices." Mathematics of the USSR-Sbornik 1, 457 – 483 (1967).
23. J. Sun, Q Qu and J. Wright "A geometric analysis of phase retrieval", Found Comput. Math, 18, 1131-1198 (2018).
24. C. Nabors, "Effects of phase errors on coherent emitter arrays", Applied Optics, vol. 33, n° 112, pp. 2284-2289, (1994).